\documentclass[referee]{aa} 

\usepackage{graphicx}
\usepackage{epstopdf}
\usepackage{txfonts}

\usepackage{hyperref}
\usepackage{gensymb}

\usepackage{xcolor}
\hypersetup{
        colorlinks   = true, 
        urlcolor     = blue, 
        linkcolor    = blue, 
        citecolor   = blue 
}

\usepackage{multirow}

\begin{document}

        \title{A type III radio burst automatic analysis system \\ and statistic results for a half solar cycle \\ with  Nan\c{c}ay Decameter Array data}

        \author{P. J. Zhang %
                \inst{1}\thanks{\email: pjer1316@mail.ustc.edu.cn}
                \and
                C. B. Wang %
                \inst{1,}\inst{2}\thanks{\email: cbwang@ustc.edu.cn, corresponding author}
                \and
                L. Ye %
                \inst{1}
        }
        \institute{
                CAS Key Laboratory of Geospace Environment, School of Earth and Space Sciences, University of Science and Technology of China, Hefei, Anhui 230026, China
                \and
                Collaborative Innovation Center of Astronautical Science and Technology, Hefei, Anhui 230026, China
        }
        
        
        
        \abstract {}
        {We design an event recognition-analysis system that can automatically detect solar type III radio burst and can mine information of the burst from the dynamic spectra observed by Nan\c{c}ay Decameter Array (NDA). We  investigate the frequency drift rate of type III bursts and the speed of electron beams responsible for the generation of the bursts.}
        {Several computer vision methods are used in this automatic analysis system. The Hough transform is performed to recognize the line segment associated with type III bursts in the dynamic spectra. A modified active contour model is used to track the backbone of the burst and estimate the frequency drift rate at different frequency channels. We run this system on the NDA data from 2012 to 2017, and give a statistical survey of the event number distribution, the starting and stopping frequencies of bursts, the frequency dependence of the drift rate, and the exciter speed using three corona density models.}
        {The median value of the average frequency drift rates is about 6.94MHz/s for 1389 simple well-isolated type III bursts detected in the frequency range 10--80 MHz of NDA observation. The frequency drift rate changes with frequency as $df/dt = -0.0672 f^{1.23}$ from a least-squares fitting. The average exciter speed is about 0.2$c$ based the density models. We do not find any significant dependence of the drift rate and the exciter speed on the solar activity cycle.}
        {}

        
        \maketitle
        %
        
        \section{Introduction}
        
        Continued research into solar radio type III burst has been made since it was first recognized and named by \citet{wild1950observations1}. The main defining characteristic of {type} III radio bursts is that their emission drifts rapidly from high to low frequency, with a drift rate of roughly 100MHz/s in the {meter wave} range. The dynamic spectra characteristics of type III bursts at meter wavelengths are described in detail and  nicely reviewed by \citet{suzuki1985bursts}, and recently by \cite{reid2014review}. It is generally believed that type III radio bursts are generated from weakly relativistic electron beams moving outward along open magnetic field lines from solar active regions. A recent statistical result reveals that all type III bursts may be associated with solar electron events from in situ observation by the WIND/3DP instrument \citep{Wang2012A}. The plasma emission mechanism is now the most commonly accepted model for the excitation of type III bursts, which was first proposed by \cite{Ginz1958onthe}, and then refined by a number of researchers \citep{sturrock1964type, melrose1980emission, robinson1998fundamental, li2008quasilinear}. The plasma emission theory is a two-step process. The electron beam first excites Langmuir waves, and then part of the energy of the Langmuir waves is converted into electromagnetic waves at either  the fundamental or the second harmonic of the local plasma frequency, or both. The plasma frequency decreases as the electron density attenuates along the radial direction from the sun. Fast electrons created by flares or jets or some other solar activities moving outward from the sun will experience {monotonic} decrease in electron density, so the observed emission frequency of the bursts decrease with time on the dynamic spectra.
        
        The frequency drift rate of type III bursts depends both on the radial velocity of the exciter electrons and on the gradient of the background electron density distribution with height, according to the plasma emission mechanism. We can use the observed frequency drift rate to estimate the speed of electron beams responsible for the generation of type III radio emissions if we have an electron density model for the solar and interplanetary density-distance scale   \citep[e.g.,][]{alexander1969type,dulk1987speeds,hoang1994interplanetary,reiner2015electron,krupar2015speed}. There is a wide range of the exciter beam speeds estimated by different researchers. The radial speed can be as low as 0.04$c$ and as high as 0.6$c$, where $c$ is the speed of light. On the other hand, if we can do {in situ} observation to get the speed and pitch angle of the fast electrons,  with a few assumptions it is possible to derive an empirical model of the electron density distribution along the trajectory of the burst source \citep[e.g.,][]{Leblanc1998Tracing}. These studies about the density and dynamic spectra complement each other, and reveal that  type III radio bursts, and  other types of solar radio emission as well, play important roles in understanding the physical processes of energy release, particle acceleration, and particle transport in the solar atmosphere and interplanetary region \citep{bastian1998radio, kontar2017imaging}.
        
        In addition, with the increase in the number of ground-based and space-based instruments for solar observations that have characteristics of high time resolution, high space resolution, and high frequency resolution, massive amounts of  scientific observation data is produced every day \citep[e.g.,][]{lemen2011interface,yan2012radio,wang2013calibration,van2013lofar}. The high quality and quantity of data is helpful in order to improve the reliability of the results from statistical analysis, to study the fine structures of solar activities, and to distinguish different theoretic models, etc. However, it is exhausting for scientists to analyze and process  data by manual operation for every frame of data. Automatic methods for  event selecting and feature finding have great significance for scientific research because of their efficiency in dealing with the massive amounts of data available. For example, a comprehensive automated feature-recognition system has been developed for the Solar Dynamics Observatory (SDO), which can automatically detect, trace, and analyze numerous phenomena; for example it can  locate the position of flare and filament, find the polarity inversion line from the magnet image, and  determine the direction of jets \citep{martens2012computer}. In solar radio physics, \cite{lobzin2009automatic,lobzin2010automatic} developed an Automated Radio Burst Identification System (ARBIS) that can detect type III and type II coronal radio bursts automatically using the solar radio spectra provided by the Radio Solar Telescope Network (RSTN) operated by the U.S. Air Force. The Radon transform and the Hough transform are chosen in their work to recognize the line segments associated with type III radio bursts and type II radio bursts on the dynamic spectrum, respectively \citep{hough1962method, duda1972use, gonzalez2012digital}. It was found that the performance of the implementation is quite high, while the occurrence probability for false positives is reasonably low. And recently, multimodal deep learning network and long short-term memory networks have been built for solar radio burst classification using the data obtained by the Solar Broadband Radio Spectrometer (SBRS) of China \citep{xu2015solar,ma2017multimodal}.
        
        In this work, inspired by the work of \cite{lobzin2009automatic}, we design an automatic recognition and information extraction system of type III solar radio bursts for dynamic spectra observed by the {Nan\c{c}ay} Decameter Array (NDA) \citep{nancay1980,nancay2000}. Several computer vision methods were employed in developing this system: 1) the Canny edge detector,  used in data preprocessing to detect and remove the bad data from instrument or environment interference \citep{canny1986computational}; 2)   the Hough transform, used to detect and recognize the line segments corresponding type III radio bursts on the spectrum image. From the recognized line segments, we can get the starting time, the starting frequency and the stopping frequency of the burst; and 3) a modified active contour method,  applied to track the spectrum backbone of a type III burst so that the frequency drift rates and intensities of the burst can be obtained at different frequency channels. The NDA data from 2012 to 2017 was chosen to perform the performance test of the system and to study the statistical characteristics of type III radio bursts.
        
        The organization of the paper is as follows. In Section 2 we briefly introduce  the data  used. In Section 3 the complete  data preprocessing, event detecting, and information extracting is described in detail. In Section 4 we present the statistical results using the information extracted from the events. In Section 5 we present the discussion and our  conclusion.

        \section{Data}
        
        We use the dynamic spectral data from NDA which observes  almost daily the Sun and Jupiter at low radio frequencies \citep{nancay1980,nancay2000}. This station was built in 1977, and has  served for more than 40 years producing spectra data for scientific use. The NDA  observes wavelengths ranging from 3 to 30 meters, corresponding to frequencies ranging from 10 to 100 MHz. It is composed of 144 helicoidal/conical antennas, 9m high and 5m  in diameter, spread across a 7000m$^2$ area. Raw data is organized by date in the form of a binary file that can be downloaded from their official site \footnote{: \href{https://realtime.obs-nancay.fr}{https://realtime.obs-nancay.fr} }. A relatively logarithmic scale is employed for the signal intensity, which takes integer values in the range of 0 to 255 in the raw data. Each file contains spectral data for one day, and an IDL procedure is provided to read the binary data on the official site. The data product has two parts, left-hand polarized and right-hand polarized, and each part has a time resolution of 1 second and a frequency resolution of 0.175MHz. In this work, we use the left-hand polarized data of solar decameter emissions in the frequency range 10--80 MHz between 2012 and 2017.
                
        \section{Method}
        The data processing pipeline starts from the raw data; the entire pipeline contains three parts: 1) preprocessing, where we first split the raw data into small segments,  subtract the slowly varying background signal, and then eliminate the bad data segments containing device calibration or lightning discharges; 2) recognition,  the core of this system, which determines whether the data segment has event of type III radio bursts or not; and 3) further analysis, used to mine more information about the radio burst from the event-active data segment.
        
        \subsection{Preprocessing}
        \subsubsection{Data segmentation}
        
        Type III bursts are fast frequency drift bursts, which can occur singly, in groups, or in storms. The typical duration of a single type III event is about 5--10 seconds in the frequency range of NDA, while the background signal intensity changes gradually on a  timescale of hours. The bright temperature of type III radio bursts can vary from $10^6$K to $10^{15}$K \citep{suzuki1985bursts} and this results in a large difference in the relative intensity of different type III events. If we analyze a data segment on a very long time span, the weak event may be ignored. For the convenience of the following processing procedures, it is necessary to split the original data into short segments.
        
        In this work, the time span of each spectra segment is 300 seconds and the segment samples every 240 seconds, which means two neighboring data segments have an overlap of 60 seconds. This is essential to make sure there is no type III event being missed in the data segmentation process. The raw data records the logarithm of the relative intensity over the 400 frequency channels in 10--80 MHz { once a second.} Thus, each data segment in this paper is a $300\times400$ two-dimensional array $I(t_i,f_j)$ of dynamic spectrum where the indices $t_i$ and $f_j$ correspond to time and frequency, respectively. The duration of 300 seconds makes it easier to identify the event. There exists slowly varying background noise which can be reduced by subtracting the average intensity at each frequency channel with the following expression:
        
        \begin{equation}
        I^\prime(t_i,f_j)=\left| I(t_i,f_j) - \dfrac{\sum\limits^{b}_{n=a} I(t_n,f_j)}{b-a}\right|.
        \end{equation}
        Here $I(t_i,f_j)$ and $I^\prime(t_i,f_j)$ denote the signal intensity before and after the noise reduction, respectively, and $a$ and $b$ are the starting and ending time index of the data segment.
        
        \subsubsection{Elimination of bad data segment }
        
\begin{figure*}
        	\centering
        	\includegraphics[width=14.3cm]{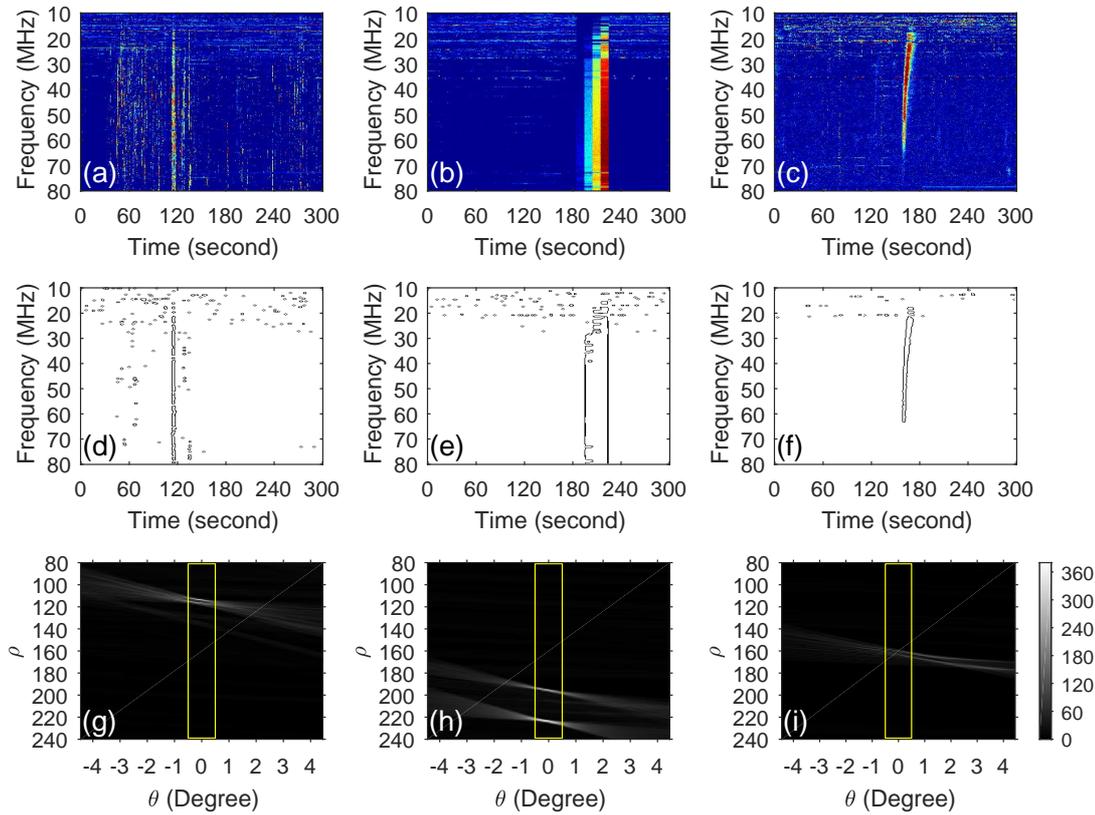}
        	\caption{From top to bottom: dynamic spectra of two bad data segments and a  type III event, their edge images detected using the Canny edge-detection method and the Hough transform results of the edge images. The left, middle, and right columns are the results of a group of lightning discharges, a calibration bar, and a normal type III burst, respectively. The type III burst was observed on Apr 22, 2013, at 14:05UT.}
        	\label{fig:1}   
        \end{figure*}

        There are a number of segments in the raw data that contain signals from device adjustment or environment interference. These types of signal satisfy the judgment of ``fast frequency drift'' in some sense, but they are not type III radio bursts. There are  two main kinds of bad data segment, as shown in Figures \ref{fig:1}(a) and \ref{fig:1}(b), respectively. One is seen as thin striped  vertical lines, which often happens in groups and may result from local lightning discharges. The other is seen as a sequence of vertical bars with clear edges, which may be the consequences of instrument calibration or local machinery interference. The common feature of these interference signals is that they contain stripes with {vertical edges as in the image}. This feature can be a great measure for us to judge whether a data segment is bad or not.

        The first step for the bad data elimination is performing edge detection using the Canny edge detector, which is widely used  \citep{canny1986computational,gonzalez2012digital}. Figure \ref{fig:1}(d) and Figure \ref{fig:1}(e) show examples of the edge detection operation applied to two kinds of bad data segments that contain a group of lightning discharges and a calibration bar, respectively. For comparison, the operation results on a normal type III burst observed on April 22, 2013, at 14:05UT is also shown in Figure \ref{fig:1}. This type III burst will be used as an example to explain the procedure of data processing. It can be seen that the edge detection operation determines the edge accurately.
        
        A standard Hough transform \citep{hough1962method} is then used to detect lines in the edge image. The Hough transform converts an image into another image in the space of Hough parameters (referred to as a Hough image in the following). Today the $\rho-\theta$ parametrization proposed by \cite{duda1972use} is in common use; it transforms an image from the $x-y$ space to the $\rho-\theta$ space using the Hesse normal form $\rho=x \cos\theta + y\sin\theta$. Here $x-y$ is the position of a point in the image or equally the row and column index of the pixel; $\rho$ is the distance from the line to the origin and $\theta$ is the angle between the line and one of the axes. In this way, objects like straight line segments in the original image are transformed into localized peaks in the $\rho-\theta$ parameter space, with longer segments producing bigger peaks. Significant peaks on the Hough image are defined as Hough peaks. Each Hough peak stands for a $\rho-\theta$ pair with which we can locate a line segment on the original image \citep{gonzalez2012digital}.
        
        The bottom three panels in Figure \ref{fig:1} display the Hough transform results of the three edge images detected from the two bad data segments and the type III burst. It is possible to  find the difference between the two Hough images transformed from bad data and that from the type III event data: the former has strong peaks near $\theta=0$, but the latter has some faint peaks slightly farther away from $\theta=0$ (generally $|\theta|>0.5\degree$ ). This occurs because the envelope of a real type III burst is more or less an oblique line due to frequency drift, while the spectrum of interference signal have nearly vertical edges. As a result, to eliminate the bad data segment, we only consider Hough peaks in the region of $\theta\in(-0.5\degree,0.5\degree)$ indicated by the yellow rectangle in Figure \ref{fig:1}, and get the line segments of these peaks. The total length of these  line segments is defined as $L_{edge}$. If $L_{edge}$ is larger than a {threshold} $L_t$ (in this work we use $L_t$ = 285), the segment is labeled as bad data and is shielded in the following operation.
        
        \subsection{Recognition}
        So far, we have subtracted the noise and eliminated the bad data  segments. It is now ready to determine whether a spectra segment contains a type III event. The recognition of type III events in a dynamic spectrum can be considered  a problem of objection recognition in the image obtained by plotting the spectrum in time-frequency coordinates.
        
        \subsubsection{Image binarization}
        Considering that the objective of this module is to recognize and locate  type III events, only the number of frequency channels in which the burst signal is visible are chosen to be important for recognition. The signal intensity is not used in this section. It is convenient to transform the grayscale image of the observed spectrum $I^\prime(t_i,f_j)$ into a binary image $B(t_i,f_j)$ , with each pixel having one of only two discrete values: one or zero. Here we use two ways of binarization, namely a threshold method and a  local-maximum method. We  find in the following subsection that the threshold method is suitable to determine the starting and stopping frequencies of a burst, while the local-maximum  method is more appropriate to estimate the average frequency drift rate of the burst.
        
\begin{figure}
        	\centering
        	\includegraphics[width=7.3cm]{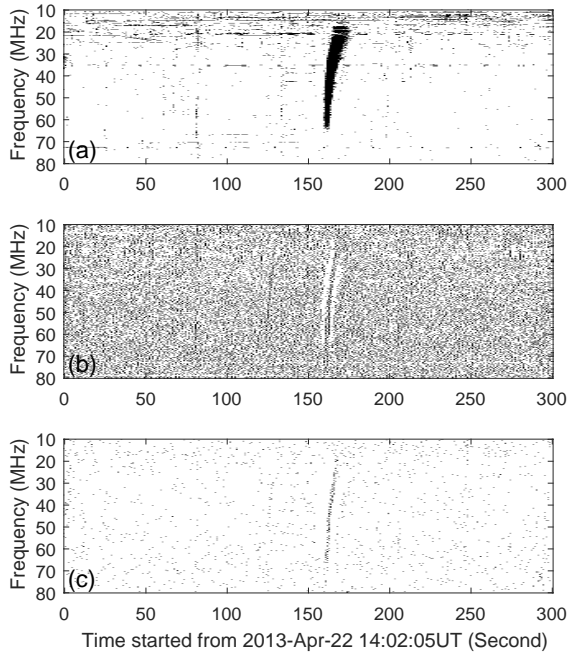}
        	\caption{Binary images of the dynamic spectrum for a type III burst using (a) the threshold method, (b) the first-order local-maximum, and (c) the the second-order local-maximum. The type III burst is the same event shown in Figure \ref{fig:1}c}
        	\label{fig:2}   
        \end{figure}

        For the application of threshold method, the key parameter is the threshold value. The signal intensities of radio bursts are different for different events so that a fixed value of threshold intensity for all data segments is not a good choice. We find that the signal intensities of type III radio burst are generally  the top 15\% of values in the intensity distribution of all pixels in a data-segment after noise reduction. Thus, we sort all pixels in each data segment  according to their signal intensities from high  to low, and set the pixel value of the first 15\% of pixels  to 1 on the binary image, while the rest of the pixels are set to a  value of 0.  Figure \ref{fig:2}(a) shows the binary image of the type III event shown in Figure \ref{fig:1}(c) using the threshold method. We can see that pixels in the event are much stronger than  in quiet area in the binary image, and they appear as a thick line segment along the event track.

        The local-maximum method is used by \cite{lobzin2009automatic} to recognize type III bursts in ARBIS using the RSTN spectra data. The pixel value $B(t_i,f_j)$ is chosen to be equal to 1 if the pixel corresponds to a local maximum with respect to time, i.e., $I^\prime(t_{i-1},f_j) < I^\prime(t_i,f_j) > I^\prime(t_{i+1},f_j)$. Performing this process to the spectra data of NDA, we find that although there is a clear black line at the location of event,  there are numerous noise points in the quiet area (see Figure  \ref{fig:2}(b)). The reason is that this algorithm is sensitive to both small and big peaks, and some signal vibration noise will result in considerable amount of local-maximum points. The rise and decay phase of a type III burst can be shown more clearly in NDA spectra than in RSTN spectra since the time resolution of NDA is higher than that of RSTN. Considering this, we propose  using the ''second-order local-maximum'' instead for binarization of the NDA spectra. In other words, the pixel value $B(t_i,f_j)=1$ if the signal intensity satisfies
                
        \begin{equation}
        I^\prime(t_{i-2},f_j)<I^\prime(t_{i-1},f_j)<I^\prime(t_i,f_j)>I^\prime(t_{i+1},f_j)>I^\prime(t_{i+2},f_j),
        \end{equation}
        otherwise $B(t_i,f_j)=0$. Using this second-order local-maximum method, we can capture the big peaks in the event center, and at the same time most of the small noise peaks are suppressed, as shown in Figure \ref{fig:2}(c). It is appropriate to use the second-order local-maximum method here. { In the following, unless specified otherwise, the term "local-maximum method'' is used for the result based on the second-order local-maximum method.  } The binary images from both the threshold method and  from the local-maximum method are used for event detection and information extraction.
        
        \subsubsection{Event detection and separation}
        Type III radio bursts have a frequency drift rate of about 5--20 MHz/s in the frequency range of NDA. The main track of a type III event has a line-like feature on the spectra image, so a standard Hough transform of the binary image is used to find features corresponding to type III bursts.
        
        \begin{figure*}
        	\centering      
        	\includegraphics[width=13cm]{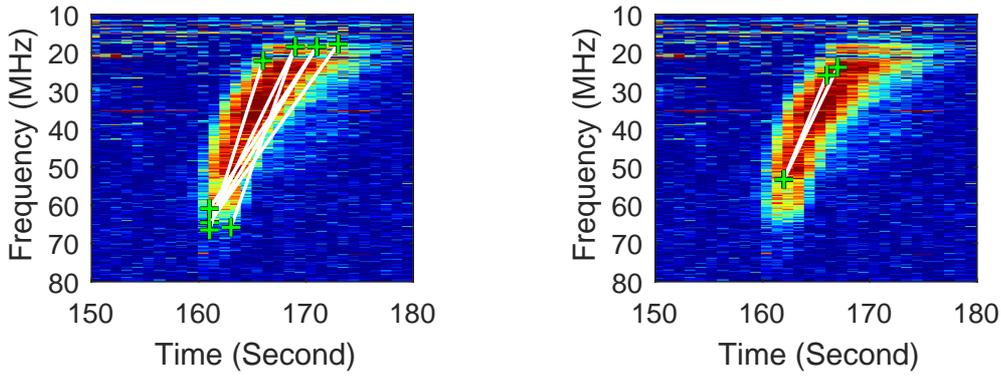}
        	\caption{Line segments detected from binary images of the threshold method (left panel) and the local-maximum method (right panel) using a Hough transform for the type III event shown in Figure \ref{fig:1}c.}
        	\label{fig:3}   
        \end{figure*}
        
        Straight segments in the original image correspond to peaks in the space of Hough parameters. If there are significant peaks in the parameter space, we can consider that there may be type III events in the spectra image. However, there may be some other type of bursts such as type II radio bursts in the data segment, whose spectra  can also  have a line-like feature on the spectra image. It is important to specify the region where the peaks are searched in the parameter space. From our experience of using the NDA data, the durations of type III bursts at a special frequency channel and in the whole frequency range of NDA are about 3--10 seconds and 5--40 seconds, respectively. Here we choose to work in the region of $\theta\in [1\degree,11\degree]$ to search for  Hough peaks associated with type III events.
        
        Figure \ref{fig:3} shows the line segments detected for the type III event on April 22, 2013, at 14:05UT, where the results from the binary images based on the threshold method and the local-maximum method are displayed in the right  and  left panel, respectively. The {criterion} for determining whether a data segment contains a type III event is as follows: if there is one or more significant peaks in the Hough image converted from the binary image of the thresh method, the data segment is determined to contain at least one event, and the threshold for judging the significance of peaks is chosen to be 265 in accumulator space in this paper. Six and two Hough peaks are detected for this event from the binary images of the threshold method and the local-maximum method, respectively. These Hough peaks are used for line segment estimation, which generate the two sets of line segments shown in Figure \ref{fig:3}.

        In some cases, a data segment may contain multiple events that need to be separated from each other. The main idea of the separation is to divide the recognized line segments into different groups according to their starting time, namely the time coordinates of their endpoints at high frequency. Line segments in the same group can be overlapping in time or separated by a short gap, but the difference between the starting time of any two line segments from different groups is greater than a threshold (10 seconds in the present work). Every group of line segments stands for an event, which can be a single burst, a fundamental-harmonic (F-H) pair burst, or a group of type III bursts with short time gap.
        
        \subsection{Information extraction}
        After a type III event is detected and separated, we get two sets of line segments for each event. The following work is to automatically determine as much information as possible to know more about the event.
        
        \subsubsection{Basic information}
        First, we can obtain several parameters directly from the line segment sets. These parameters include the starting and stopping frequencies of the event, the occurrence time of the event, and the average frequency drift rate of the event.
        
        We can see from Figure \ref{fig:3} that using the thresh  method for binarization, the endpoints of line segments lay more closely on the starting and stopping frequency of the type III event, while the line segments align more closely to the backbone of the event track using the local-maximum method. Thus, the starting frequency ($f_{start}$), the stopping frequency ($f_{stop}$), and the occurrence time ($t_{start}$) of an event are chosen to be the maximum  and minimum value of the frequencies, and the earliest value of the time given by the endpoints of line segments detected using the thresh method, respectively. The average slope of the line segments from the local-maximum method gives the average frequency drift rate ($v_f$) of the event. For the event on April 22, 2013, at 14:05 UT, the parameters $f_{start}$, $f_{end}$, $t_{start}$, and $v_f$ are estimated to be  64.1 MHz, 19.5 MHz, 14:04:46 UT, and -5.9 MHz/s, respectively.

        \subsubsection{Frequency drift rate at different frequency}
        For even more information, such as the signal intensity distribution along the radio burst and the frequency drift rate at different channels, {we need to track the backbone of the event}. The tracking method adopted here is a modified active contour model. The active contour, also called a snake, is a framework in computer vision for delineating an object outline from a possibly noisy 2D image \citep{kass1988snakes,pal1993review}. A snake is an energy minimizing, deformable spline influenced by constraint and image forces that pull it towards object contours and internal forces that resist deformation. To leave the spline convergent at the backbone of the event in this work, the expression of the force applied to a node of the snake is chosen to be
        
        \begin{equation}
        F_{total} =  \alpha \nabla I(x,y) + \beta \kappa -\gamma \vec{v},
        \label{eq:total}
        \end{equation}
        where $I(x,y)$ is the pixel intensity of the node at position $(x, y)$; $\kappa$ denotes the curvature of the line at the node; $\vec{v}$ is the velocity of the node; and $\alpha$, $\beta$, and $\gamma$ are three positive coefficients.  The first term on the right side of Equation (\ref{eq:total}) stands for the gradient force from the image, which lets the node move toward the event backbone. The second term denotes the bending force at the node, which can keep the line straighter to avoid the spline being guided by noise data too much. The third term is the damping force, which will make the line stay still at the balance position. Here, we assume that the damping force is large enough such that at any time the acceleration of every node is almost zero. The spline is approximately at mechanical equilibrium state, so from Equation (\ref{eq:total}) we can deduce the motion equation of every node as follows:
        
        \begin{equation}
        \vec{v} \approx (\alpha \nabla I(x,y) + \beta \kappa )/\gamma.
        \label{eq:motion}
        \end{equation}
        
\begin{figure}
        	\centering              
        	\includegraphics[width=5.5cm]{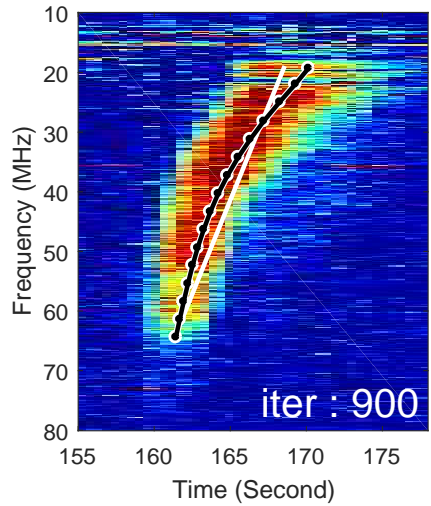}             
        	\caption{Backbone tracked by modified active contour model for a type III burst. The white line and the black chain stand for the pre-estimated line and the traced backbone at the final iteration, respectively.}
        	\label{fig:4}   
        \end{figure}

        The iteration process for the deformation of the spline is as follows. We first generate 16 uniformly spaced nodes on an initial line segment whose slope and endpoints are given by the average frequency drift rate, and the starting and stopping frequencies as estimated in the above subsection. Then we apply Equation (\ref{eq:motion}) to every node at each step, and iterate the motion equation until the spline aligns closely to the backbone of the event (here we use 900 iterations in practice). The black chain in Figure \ref{fig:4} shows the final backbone found for the type III burst on April 22, 2013, at 14:05UT.
        
        After the operation of the modified active contour model, we have a curve that {represents} the trend of the event consisting of 16 nodes shown as black points in Figure \ref{fig:4}. The frequency drift rate $df/dt$ between two neighboring nodes can be estimated by their time and frequency differences. In addition, we  note that the modified active contour model works fine for clean and strong events, but it may fail to  accurately find the backbone of an event in some cases. For example, the backbone may be misled by strong noise at low frequencies or an accompanied component in an F-H pair event, and the interruption of the dynamic spectrum can cause the  backbone to be incomplete.
        
        \section{Statistical result}
        
        With this automatic recognition and information extraction tool we can perform some quantitative studies on the statistical behaviors of type III radio bursts. The NDA data from 2012 to 2017 was chosen to perform the performance test of the system, which covers about a half solar activity cycle. The entire data set is divided into 438840 data segments, about 240 segments per day  most of the time, except during the period from July 2016 to January 2017 and from June 2017 to December 2017 due to data gap. We  found that 4250 data segments contain type III events based on the recognition system. Here an event may be a single burst, an F-H pair burst, or a group of type III bursts with short time gap, as we  mention above. Moreover, in order to study the frequency dependence of the frequency drift rate and the exciter electron beam speed of type III radio burst, we manually select 1389 simple well-isolated type III events whose backbones are clear and have been tracked accurately.
        
        \begin{figure}
        	\centering
        	\includegraphics[width=6cm]{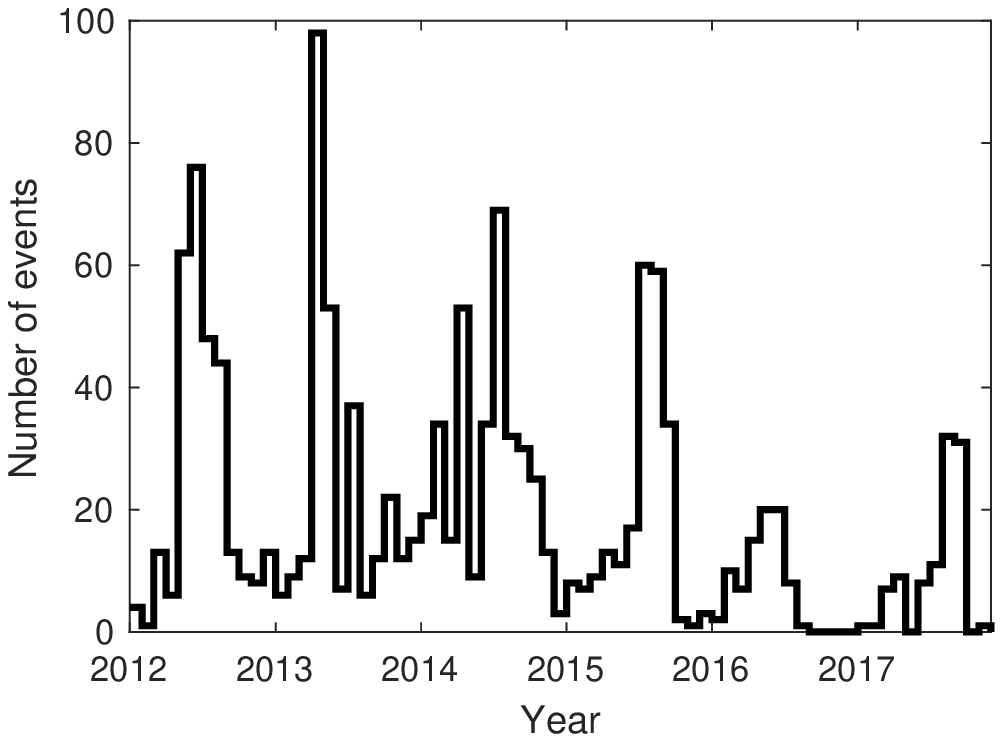}
        	\includegraphics[width=6cm]{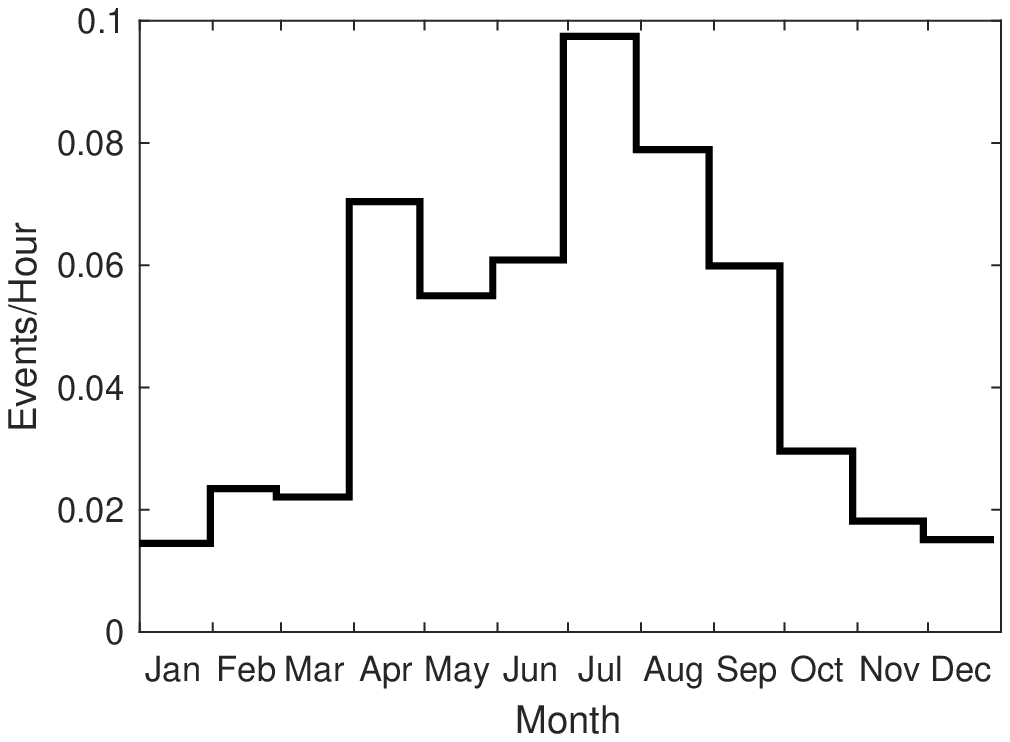}
        	\includegraphics[width=6cm]{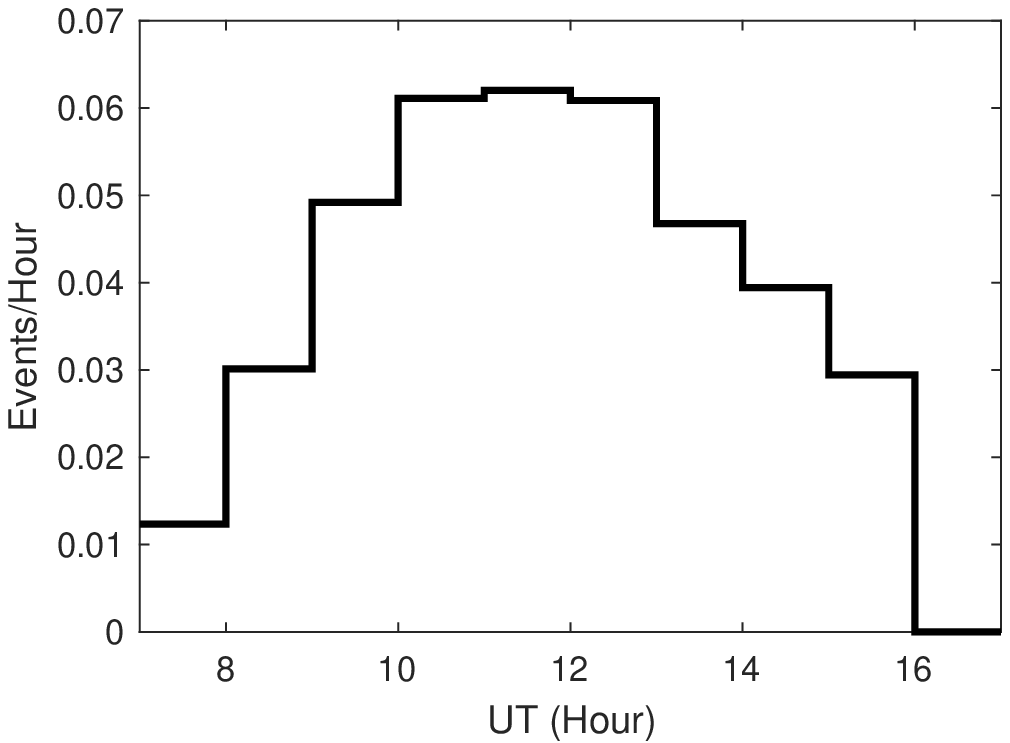}             
        	\caption{Histograms of the number of events in every month in different years (top panel), the monthly accumulated event occurrence rate (middle panel), and the hourly accumulated event occurrence rate (bottom panel) for 1389 simple well-isolated events. The occurrence rates are normalized to observation time by dividing the number of events by the total observing hours.}
        	\label{fig:5}   
        \end{figure}
        
        The 4250 data segments with type III events and the 1389 simple events are similar in temporal distribution. Figure \ref{fig:5} displays the histograms of 1389 simple events. More events can be found at higher solar activity (from 2012 to 2014) than at lower activity (from 2016 to 2017). The extremely low number between the end of 2016 and the beginning of 2017 is due to the data gap. The most interesting result is the seasonal variation in the event occurrence rate, as shown in the middle panel of Figure \ref{fig:5} where the normalization of histogram to observation time is applied to exclude possible seasonal change of the observation time. The event occurrence rate between April and September is significant greater than that between October and March in a year. We do not find any similar seasonal variation in the number of sunspot during these years, so this seasonal variation of events is not due to the change in solar activity. We  checked the NDA data month by month with naked eyes, and found that NDA really observed more type III bursts in summer than in winter.  We suggest that this seasonal variation may be due to the change in solar zenith angle in summer and winter. Considering that the zenith angle also varies during the observation time of a day, we  surveyed the event occurrence rate at different hours in a day (see bottom panel of Figure \ref{fig:5}). These results indicate that more type III events can be observed by NDA when the sun is at smaller zenith angle (near noon or summer) and vice versa. The observing frequencies of NDA are close to the cutoff frequency of the ionosphere. When the zenith angle is larger, the solar radio waves are incident on the stratified ionosphere more obliquely. It will be harder for the radio burst signal to penetrate through the ionosphere due to wave refraction and absorption, so fewer type III events can be observed for larger zenith angle. This is similar to the propagation of communication radio waves in the ionosphere, as demonstrated by ray tracing. The rays with large elevation angles can escape from the ionosphere, while the rays with small elevation angles will be reflected back \citep[e.g.,][]{ huang2006real}.

        \begin{figure}
        	\centering
        	\includegraphics[width=6.5cm]{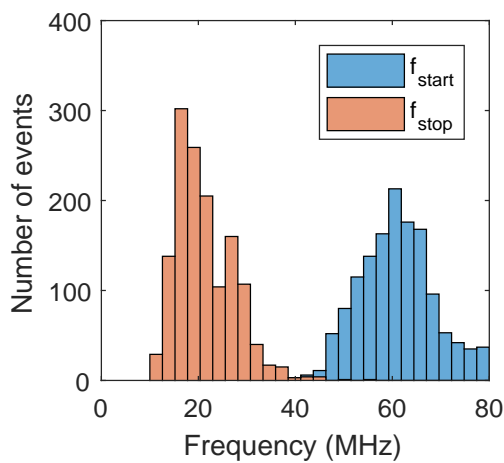}
        	\caption{Histograms of the starting and stopping frequencies for 1389 type III bursts.}
        	\label{fig:6}
        \end{figure}

        The starting and stopping frequency of the event can show the height range where the radio wave can be excited, or from where the wave can propagate to the earth and be observed by NDA. Figure \ref{fig:6} shows the histogram of the starting and stopping frequencies of the events. It is found that most of the events have a stopping frequency near 20MHz and a starting frequency near 60MHz, with a median value of  21.3MHz and 60.9MHz, respectively. However, it is important to note that the starting and stopping frequency obtained here may not represent the actual emission frequency range of a special type III burst, due to the limited frequency range of NDA observation and the continuous noise at low frequencies of 10--25MHz.

        \begin{figure}
        	\centering
        	\includegraphics[width=7.8cm]{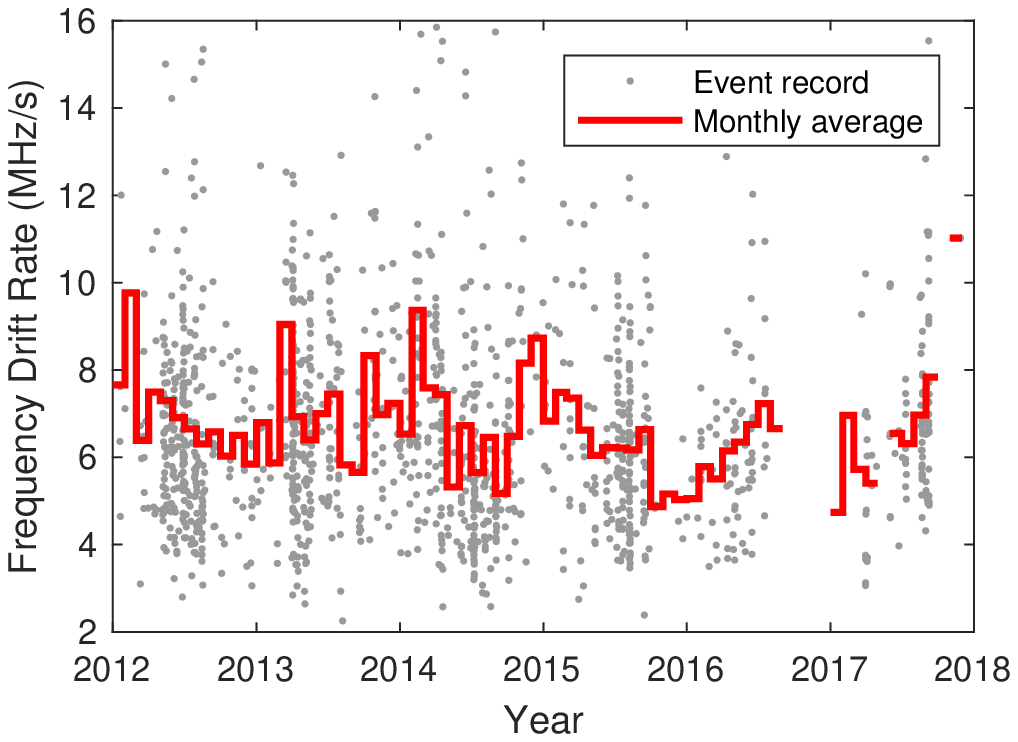}
        	\includegraphics[width=7.3cm]{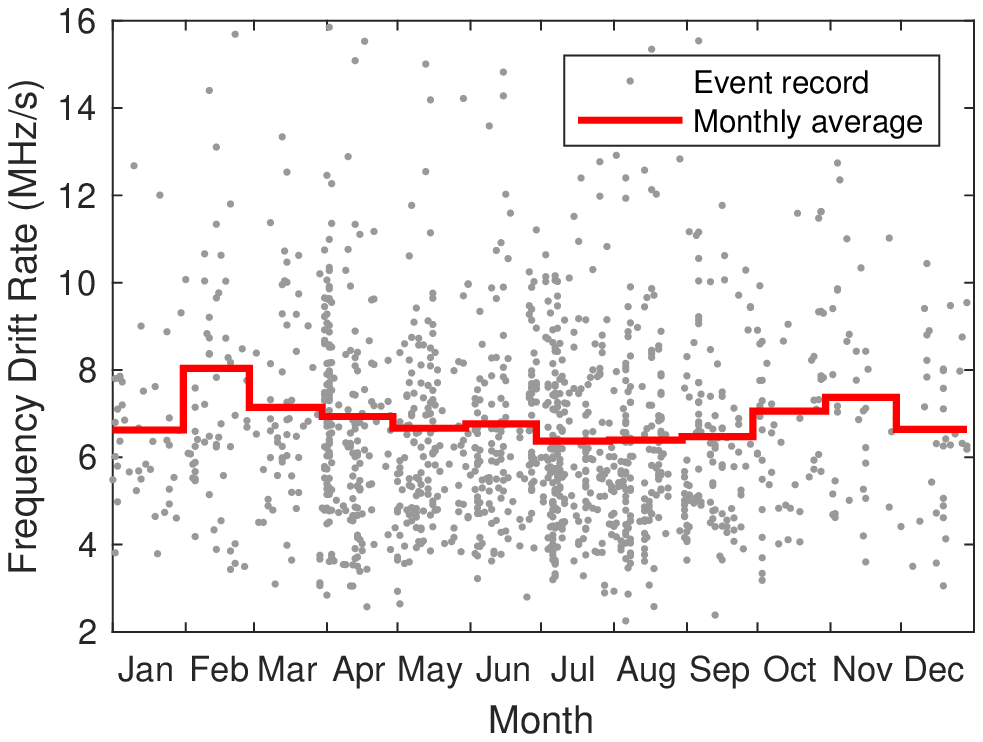}
        	\caption{Scatter plots and temporal variation of the frequency drift rate;  each gray point represents the estimated average drift rate of a type III burst. The red lines in the top and bottom panels indicate the monthly average value in different years and the monthly accumulated average value of all years, respectively.}       
        	\label{fig:7}   
        \end{figure}
    
        Figure \ref{fig:7} shows the frequency drift rate scatter-plots of the 1389 well-isolated type III events and the variation in their monthly averaged values with time. We note that the drift rate shown here is the average drift rate of an event, which is extracted directly from the slope of line segments using the local-maximum method  described  above. The absolute values of the average frequency drift rate of type III events mainly spread over the range from  2 MHz/s to 16 MHz/s, with a median value of 6.94 MHz/s. We do not find any significant difference between the drift rates measured at different solar activities. The drift rate is slightly lower near summer than winter, but this may be caused by the different number of events used for statistics. Many more events are recognized near the summer in a year based on our recognition system. A table where we list the observation time, starting/stopping frequencies, and average frequency drift rates of the 1389 type III bursts is provided through the Centre de
Donn\'{e}es astronomiques de Strasbourg (CDS).

		\begin{figure}
			\centering
			\includegraphics[width=8cm]{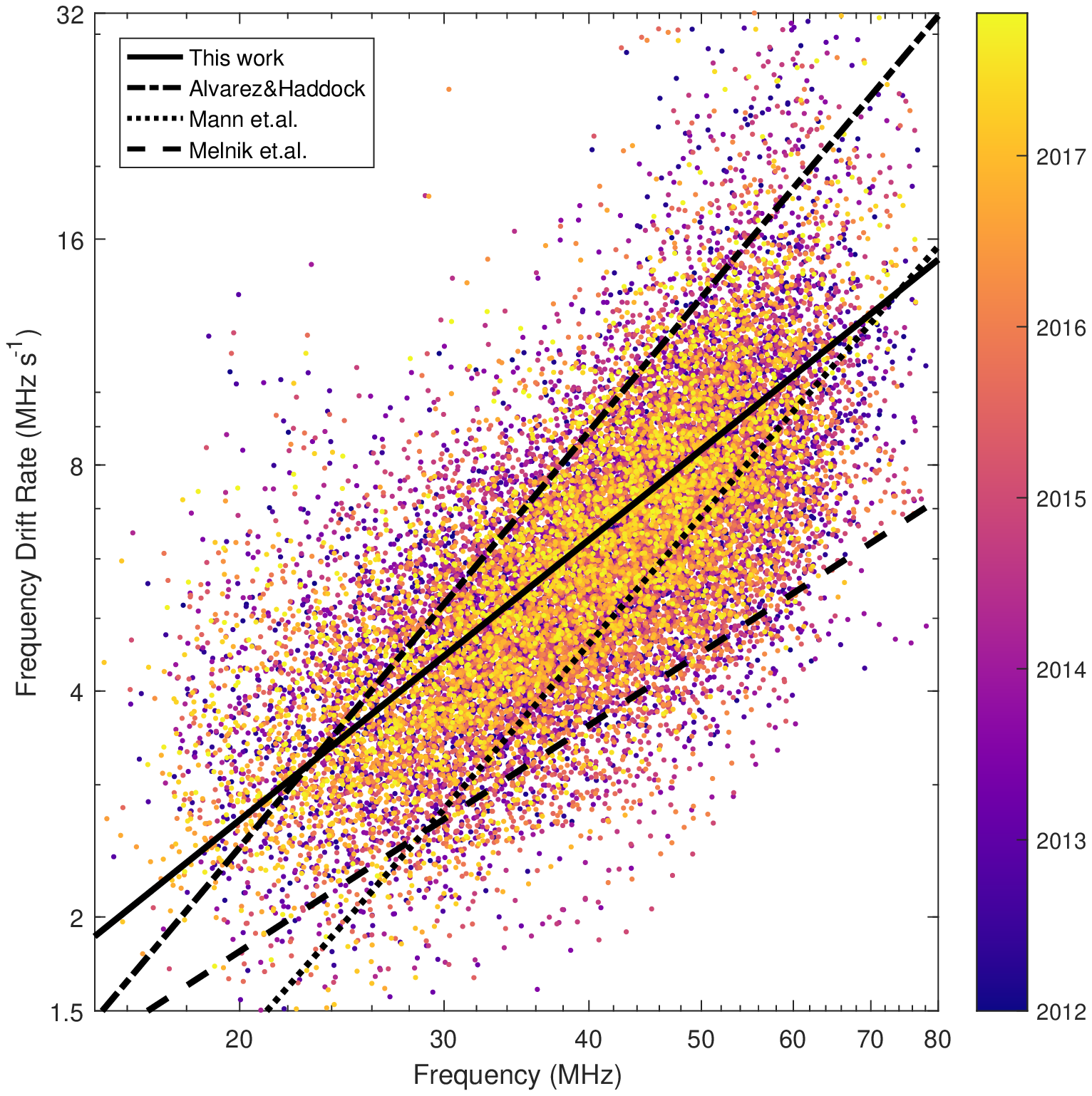}
			\caption{Scatter plot of the frequency drift rate vs frequency for 1389 simple type III bursts;  the data points are color-coded according to the occurrence time of the corresponding event. The solid line indicate the least-squares fitting result $df/dt=0.0672f^{1.23}$ in this work. The dot-dashed, dotted, and dashed lines stand for the results in the literature obtained by \citet{Alvarez1973Decay}, \citet{Mann1999A}, and \citet{melnik2011observations}, respectively}
			\label{fig:8}   
		\end{figure}

		\begin{figure*}
			\centering
			\includegraphics[width=6.5cm]{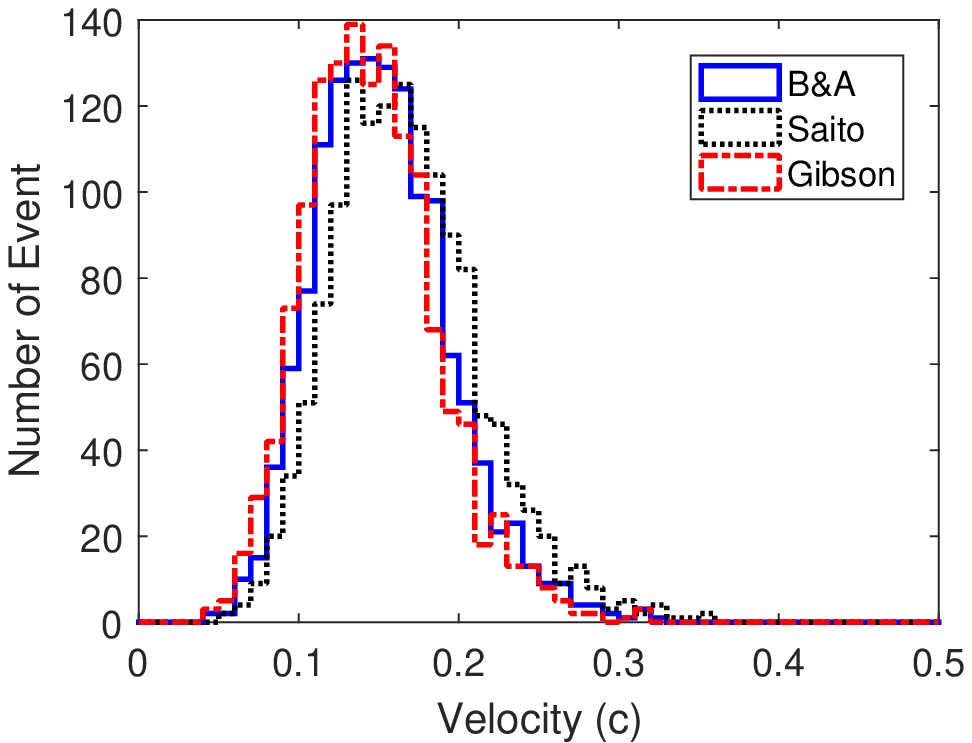}
			\includegraphics[width=6.5cm]{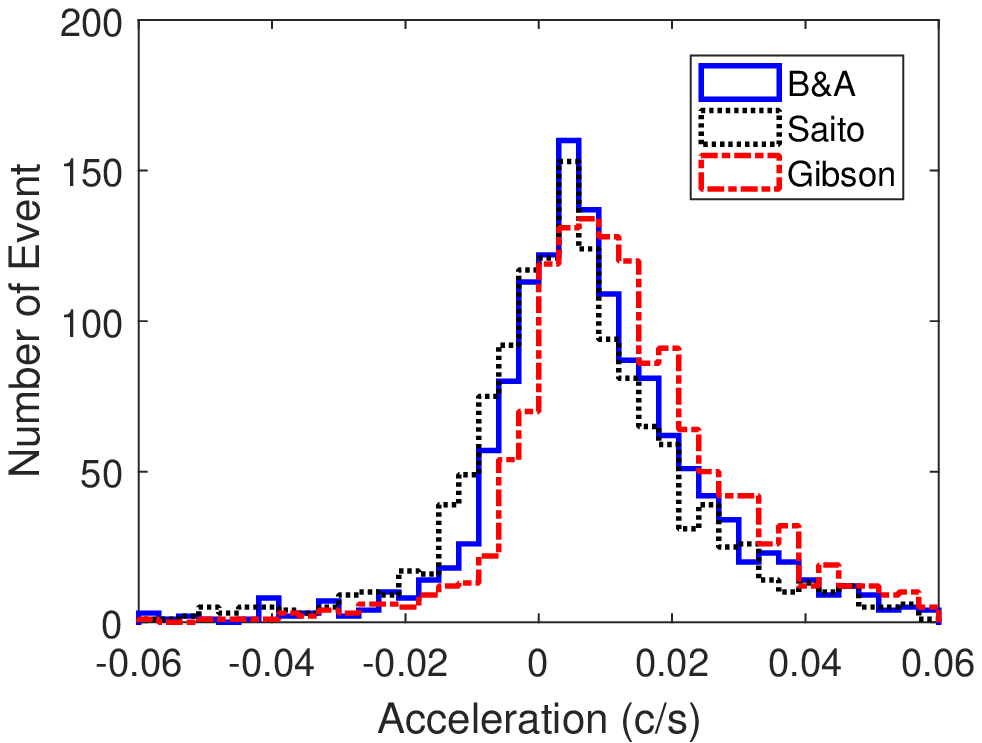}    
			\includegraphics[width=6.5cm]{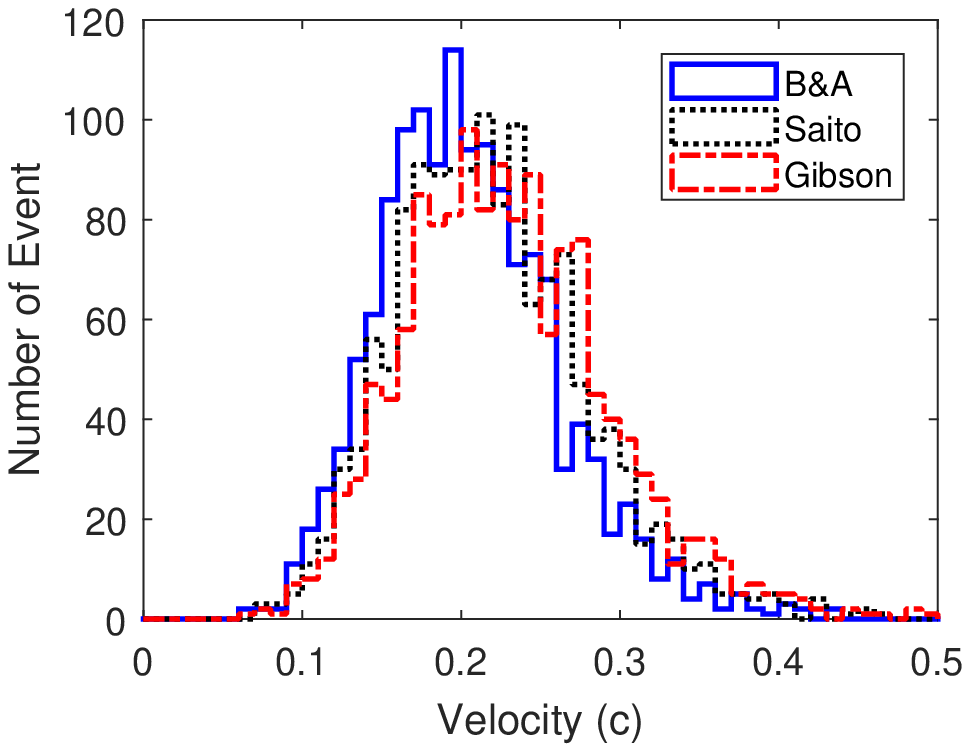}
			\includegraphics[width=6.5cm]{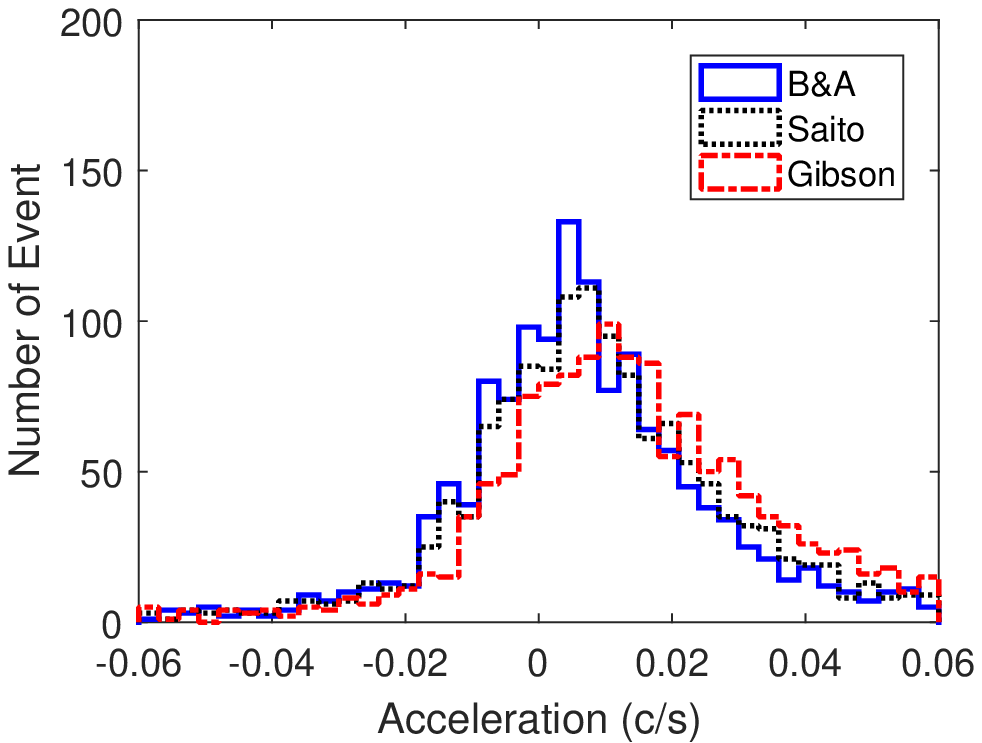}            
			\caption{Distribution of the radio source linear velocity and acceleration derived from three different electron density models of \citet{allen1947interpretation}, \citet{Saito1977A}, and \citet{gibson1999solar}. The top and bottom panels are the results based on the assumptions of fundamental and harmonic wave emission, respectively.}
			\label{fig:9}   
		\end{figure*}
	
        Using the modified active contour model, we can track the backbone $f(t)$ of an event and estimate its frequency drift rates $df/dt$ at different frequency $f$. The drift rates generally decrease with decreasing frequency in an event. Figure \ref{fig:8} shows the frequency drift rates at different observation frequencies for the 1389 simple events in logarithmic scale, where the data points are color-coded according to their observation time. We did not find significant differences in frequency drift rates for the different years. The solid line represents the least-squares fits to the data using a target function of the form $|df/dt|=a f^b$  .The best fitted values are $a=0.067\pm0.006$ and $b=1.23\pm0.02$ with 95\% confidence. For comparison, several fitting results in the literature on the frequency dependence of drift rate are also presented in Figure \ref{fig:8}, which are based on observations in different frequency ranges performed by different authors.

        According to the plasma emission mechanism, the observed frequency of the type III burst is the local plasma frequency or its harmonic at the source region. The drift rate of type III bursts can be used to estimate the speed of the exiting electron beam. Assuming a coronal density model to give electron density as a function of distance $n_e(r)$, we can find frequency as a function of distance $f(r)$. Then for each event the distance-time relation $r(t)$ about the movement of the exiting beam can be obtained from the tracked backbone $f(t)$. We can derive the radial velocity of the exciter electrons by linearly fitting the distance-time relation, and the acceleration by the second-order fitting. Figure \ref{fig:9} shows the distribution of the linear velocity (left panels) and the acceleration (right panels) deduced from the 1389 simple events based on three different density models. {Most of the 1389 events are structureless (no F-H structure,  which is generally believed to be the harmonic component) type III bursts \citep{dulk1980position, suzuki1985bursts}. Both fundamental wave and harmonic wave are assumed for the observed emissions in this survey for comparison, when we apply the density model to convert frequencies to radial distances.} The results based on the fundamental and harmonic assumptions are displayed in the top and bottom panels in Figure \ref{fig:9}, respectively. The solid lines, dotted lines, and dashed lines represent the results using three density models derived by \cite{allen1947interpretation}, \cite{Saito1977A}, and \cite{gibson1999solar}, respectively. Our preference for these models lies in the fact that they were derived completely independently of any radio data. The density model given by \cite{baumbach1937strahlung} and \cite{allen1947interpretation} is known as the Baumbach--Allen density model, which is derived from the coronal photometric data. \cite{Saito1977A} used white light coronagraph data from Skylab to derive the corona density model. \cite{gibson1999solar} used the method of a Van de Hulst inversion to derive the corona electron density distribution from corona observations of both visible white light and extreme ultraviolet (EUV) emission. The mean speed of the exciter deduced using the three density models are about 0.16$c$ and 0.22$c$ for the fundamental and harmonic assumptions, respectively. The acceleration has a Gaussian distribution with small positive mean and large standard deviation for all three density models.

        \section{Conclusions and discussion}
        An event recognition-analysis system based on computer vision methods is designed in this work, which can automatically detect type III solar radio bursts and can  mine information of the bursts using the dynamic spectra observed by NDA.
        
        The system is composed of three parts. The first part is the preprocessing module. This module includes splitting the raw data into short pieces, reducing the noise, and eliminating the ``bad data'' segments by using the Canny edge detection and Hough line detection methods. The second part is the type III event recognition module, which is inspired by the work of \cite{lobzin2009automatic}. However, the time resolution of the NDA spectrum is higher than that of the RSTN data, and the duration of a type III burst is generally longer in the NDA observation frequency range than that in the frequency range of RSTN. Therefore, both a second-order local-maximum method and a thresh method are adopted for the image binarization of dynamic spectrum in present work instead of the first-order local-maximum method used by \cite{lobzin2009automatic}. Then, a standard Hough transform of the binary image is performed to find the straight line features associated with type III bursts. If there are significant Hough peaks in the Hough image, the data segment is considered to contain type III events. The third part is the information extraction module. In this module, several basic parameters about the event, such as its occurrence time, starting and stopping frequencies, and average frequency drift rate, can be estimated directly from the line segments corresponding to the Hough peaks detected. Moreover, a modified active contour model is used to track the backbone of the event spectrum by iteration, and the frequency drift rates at different frequency can be obtained from the backbone curve.
        
        Applying this system to the NDA data from 2012 to 2017, 4250 data segments are detected to contain type III events, from which we select 1389 single well-isolated bursts whose backbones are well tracked using the modified active contour model. 
As expected, more events are recognized in years of high solar activity. An interesting finding is that there is a seasonal variation of type III events, namely, the number of events observed between April and October are significantly greater than those in other months in a year. We consider that this seasonal variation may be due to the change of zenith angle of the sun in different seasons.
        
        The median values of the starting frequency and the stopping frequency of type III bursts are 60.9MHz and 21.3MHz in the observation frequency range of NDA, respectively. However, the frequency extent of type III bursts covers a wide range. Both the starting frequency and the stopping frequency can be markedly different from burst to burst, varying from hundreds of MHz to tens of kHz \citep{malville1962characteristics, cane2002solar, Leblanc1995, kruspar2014}. It has been found that there is an anti-correlation between type III starting frequencies and hard X-ray spectral indices in time during solar flares \citep[e.g.,][]{Kane1981} which can be used to deduced the acceleration height and size of the electron beam \citep{reid2011characteristics, reid2014low}. What process causes the stopping frequency is still unclear. The recent simulation results by \citet{reid2015stopping} indicate that type III stopping frequency can be influenced by a number of effects such as the expansion rate of the magnetic flux tube guiding the electron beam, the injected beam density, the energy spectral index of the beam, and the level of large-scale density fluctuations in the background solar atmosphere, {all of which are} consistent with the postulations by \citet{Leblanc1995}. 
        
        We do not find any dependence of the drift rate on the solar activity cycle, indicating that there may be no systematic difference between the height scale of corona electron density in source regions of type III bursts at solar maximum and that at solar minimum, or for the speed of electron beams producing type III bursts. The drift rate changes with frequency as $df/dt=-0.067f^{1.23}$ from a least-squares fitting in the frequency range 10--80MHz of NDA. As shown in Figure \ref{fig:8}, most of the scattered data points are located below the line representing the expression of $df/dt=-0.01f^{1.84}$, which was obtained by \cite{alvarez1973solar} {who} used the rise time of type III bursts between 3.5 MHz and 50 kHz and combined them with eight other studies up to 550 MHz. Of course, we should  note that the drift rate of backbone tracked in the present study is the drift rate of the peak flux. The drift rate of a type III burst onset is generally less than the drift rate of its peak flux since the rise time of type III burst increases with the decreasing of frequency  \citep[e.g.,][]{Alvarez1973Decay}. \cite{melnik2011observations} found that the drift rate of peak flux  linearly increased with frequency ($df/dt \approx f$) between the frequencies 30 MHz to 10 MHz for a number of powerful radio bursts during July - August in the solar maximum of 2002. Their results are smaller than most of the observation values in this paper as shown in Figure \ref{fig:8}. In addition, \cite{Mann1999A} also reported a relationship of $df/dt = -0.007f^{1.76}$ for the coronal and associated interplanetary (IP) type III bursts on December 27, 1994, in the frequency range from 40 kHz to 85 MHz.
        
        The drift rate of type III bursts can be used to estimate the electron exciter speeds which have attracted much attention in the literature. Earlier studies implied that the speed of the exciter should be weakly relativistic and the average radial velocity is approximately 0.35$c$ \citep{wild1950observations1, alexander1969type, fainberg1970type} . \cite{dulk1987speeds} investigated 28 IP type III bursts accompanied with in situ measurement of Langmuir waves by the ISEE-3 spacecraft. They found that the range of exciter speed is 0.07--0.25$c$ with an average speed of 0.14$c$, which is considerably lower than the earlier accepted values of 0.3--0.5$c$. There is no evidence in their data to show a systematic increase or decrease in exciting electron speed with distance from the Sun to the Earth orbit. However, more recent studies about the trajectory of the type III burst exciters implied the exciter electrons experiencing a deceleration in the solar wind, based on the observation of WIND/Waves and STEREO/Waves instruments \citep{krupar2015speed, reiner2015electron}. The average value of the exciter speed obtained by \cite{krupar2015speed} is 0.1$c$, ranging from 0.02$c$ to 0.35$c$ for 29 IP type III events, but the average exciter speeds deduced by \cite{reiner2015electron} ranged from 0.17$c$ to 0.35$c$ for five IP type III bursts associated with local radio emission near the spacecraft.
        
        In the present work, we also derived the linear velocity and acceleration of the exciter electrons for 1389 isolated bursts using three corona density models \citep{allen1947interpretation,Saito1977A,gibson1999solar}. These density models are all derived from photometric data of white light or EUV corona observation, being independent of any radio data. The linear velocities distribute in the range of 0.05--0.5$c$ and 0.04--0.4$c$, with a mean value of about 0.16$c$ and 0.22$c$ for the fundamental and harmonic assumptions of observed radio emissions, respectively. The sources of radio emission at 10 MHz to 80 MHz are located in the height from 1.1 to 3.1 solar radii according to these density models. Our results of the speed near the Sun are generally consistent with the statistical values obtained by \cite{dulk1987speeds} and \cite{krupar2015speed} in the solar wind. The median values of the acceleration distribution are positive for three density models, though the standard deviations are large. This seems to suggest that the exciter electrons experience weak acceleration in the corona (between 80 MHz and 10 MHz). One possible mechanism for the acceleration is that nonthermal electrons injected into the open magnetic field lines from the flare reconnection site have a ring-beam velocity distribution in phase space. When these nonthermal electrons propagate outward from the Sun, the momentum of the ring component that is perpendicular to the field line will transfer to the parallel beam component, supposing the conservation of energy and magnetic momentum. This will produce an increase in the exciter's parallel velocity along the magnetic field line. As mentioned above, the results by \cite{ krupar2015speed} and \cite{reiner2015electron} show that the electron beam decelerates in the interplanetary range (below 1 MHz). The acceleration and deceleration in different height ranges seem to indicate that the beam speed reaches its maximum between 10 MHz and 1 MHz, which corresponds roughly to the critical point of the Parker solar wind model. This may be observed by the Parker Solar Probe \citep{fox2016solar}.
        
        Finally, we would like to point out that the plasma emission mechanism is assumed in this work for the convenience of comparison with previous works on the exciter speeds of type III bursts. Recently, a direct emission mechanism based on the electron cyclotron maser (ECM) emission mechanism has been proposed for type III solar radio bursts \citep{wu2002generation,yoon2002generation,wu2005altitude,wang2015scenario,chen2017self}. If we consider the ECM emission mechanism, the source speed derived above  corresponds to the speed of the apparent source.

        \begin{acknowledgements}
                The Nan\c{c}ay Radio Observatory / Scientific Unit of Nan\c{c}ay of the Paris Observatory (USR 704-CNRS, supported by University of Orleans, OSUC, and Center Region in France) providing access to NDA observations accessible online at \href{http : //www.obs-nancay.fr}{http : //www.obs-nancay.fr}. The research was supported by the National Nature Science Foundation of China (41574167 and 41421063) and the Fundamental Research Funds for the Central Universities (WK2080000077). We thank the anonymous referee for the helpful comments on the manuscript.
        \end{acknowledgements}
        
        %
        %
        \bibliographystyle{aa}
        \bibliography{cite}

\end{document}